# Impact of DER Communication Delay in AGC: Cyber-Physical Dynamic Simulation

Wenbo Wang, Xin Fang, Anthony Florita

National Renewable Energy Labortory, Golden, CO, 80401, US

*Abstract*— Distributed energy resource (DER) frequency regulations are promising technologies for future grid operation. Unlike conventional generators, DERs might require open communication networks to exchange signals with control centers and their aggregators, possibly through DER aggregators; therefore, the impacts of the communication variations on the system stability need to be investigated. This paper develops a cyber-physical dynamic simulation model based on the Hierarchical Engine for Large-Scale Co-Simulation (HELICS) to evaluate the impact of the communication variations, such as delays in DER frequency regulations. The feasible delay range can be obtained under different parameter settings. The results show that the risk of instability generally increases with the communication delay.

## I. INTRODUCTION

With the rapid deployment of distributed energy resources (DERs), their capability to provide grid services such as frequency regulation is being investigated [1]. Unlike conventional generators, which use a dedicated communication channel to provide automatic generation control (AGC) [2], DERs might require open communication networks to exchange control signals with system control centers, possibly through DER aggregators. The open networks expose several vulnerabilities of the DER AGC services, such as extended communication latency, increased packet loss, and cyberattacks (e.g., false data injection); therefore, it is imperative to study the impact of the communication variations in DER AGC on the system frequency stability to ensure reliable grid operation.

Several existing works in the literature on the communication delay in the system load frequency control (LFC) focus on conventional generation providing frequency regulation services [3]–[5]. In these studies, the communication delay margins were evaluated analytically based on Lyaponuv stability theory and relatively small systems; however, they did not capture the discrete nature of secondary frequency regulation (i.e., AGC). From an analytical point of view (continuous, nonlinear control models), the previous delay evaluation methodology might not be well suited for DER AGC control analysis with discrete control signal.

To study the impact of the communication delay on DER AGC and system frequency stability, this paper proposes a cyber-physical dynamic simulation (CPDS) model. The discrete AGC control signal is sent from the system control center to DER aggregators every 4 seconds in a discrete manner. The DER dynamic response and system frequency response are modeled in ANDES, which is an open-source, dynamic simulation tool. The whole CPDS cosimulation model is developed using the Hierarchical Engine for Large-Scale Co-Simulation (HELICS). Then, the feasible space of the communication delay is obtained through multiple simulations with the proposed model.

## II. DER DYNAMIC MODEL INCLUDING LFC MODEL WITH COMMUNICATION DELAYS

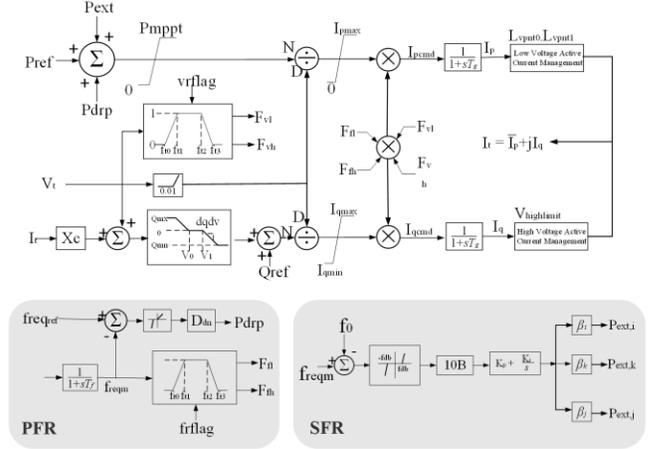

Fig. 1. DER frequency dyanmic model, PFR, and SFR.

### A. DER Frequency Dynamic Model

Here we use the Western Electricity Coordinating Council PVD1 model [6] to represent DER frequency dynamic behavior, as shown in Fig. 1. Generic models of primary frequency response (PFR) and secondary frequency response (SFR) for DERs are also included in the figure. Note that a variable representing the limit of maximum available power from maximum power point tracking (MPPT), $P_{mppt}$, has been added to the dynamic model. This allows the model to consider the photovoltaic production headroom, or other user-defined limits, etc. More details of the dynamic model can be found in [6].

*1) PFR*

PFR uses droop control: when the frequency drop is larger than a PFR deadband, the DER changes its active power output accordingly. An additional power output, $P_{drp}$, is added to the generation output:

$$P_{drp} = \begin{cases} \frac{(60 - db_{UF}) - f}{60} D_{dn}, f < 60 \\ \frac{f - (60 + db_{OF})}{60} D_{dn}, f > 60 \end{cases} \quad (1)$$

where $db_{UF}$ and $db_{OF}$ are the underfrequency and overfrequency deadband; and $D_{dn}$ is the per-unit power output change to 1-p.u. frequency change (frequency droop gain).

*2) SFR*

SFR is enabled by an AGC model that includes two components: an area-level estimation of the area control error (ACE) from (2) [7] and a plant-level control logic that receives the ACE signal and sets the reference power, $P_{ext}$, for each plant. For simplicity, it is assumed that there is one area in the simulation and no interchange with other areas.

$$ACE_{tt} = 10B(f_{reqm,tt} - f_0) \quad (2)$$

where $tt$ is the AGC time interval index; $ACE_{tt}$ is the ACE at the AGC interval $tt$; $f_{reqm,tt}$ is the measured system frequency at the AGC interval $tt$; $f_0$ is the system reference frequency (60 Hz); $B$ is the frequency bias in MW/0.1 Hz; and $f_{db}$ is the frequency error tolerance deadband.

As shown in Fig. 1 for the SFR, the proportional integral (PI) control is applied to the ACE signal; $K_p$ and $K_i$ are the coefficients of the PI controller. The ACE signals are updated every 4 seconds to represent their discrete nature in the field. The output from the PI controller is then passed on to each AGC generator considering the unit's participation factor ($\beta_i$ is the $i$-th unit's participation factor), resulting in the final control reference $P_{ext,i}$. Note that the participation factor of each unit is decided by a real-time economic dispatch normally updated every 5 minutes.

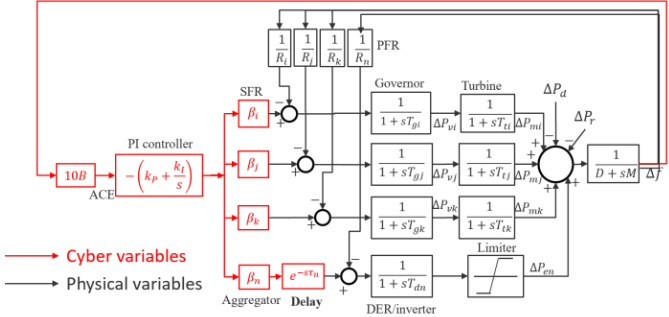

Fig. 2. Schematic model of cyber (red) and physical in transmission (black) with delays.

### B. LFC with Delays

The delays in LFC come from the fact that the open network (e.g., mobile or fixed broadband) might be required for DER to provide frequency services. The block diagram shown in Fig. 2 combines PFR and SFR with added delay blocks. Red represents the cyber variables where communications are required, whereas black represents physical variables (governors, turbines, inverters) and locally controlled PFR.

### III. CPDS COSIMULATION

This section develops the proposed CPDS cosimulation for studying delay impacts in DER AGC. This cosimulation is based on the HELICS platform and the open source power system simulator ANDES. HELICS is an open-source, cyber-physical cosimulation framework for energy systems. A few key concepts of HELICS that are relevant here: *federates*, *broker*, *simulators*, *messages*. For more details, see [8].

Assume that the overall system comprises a transmission system, a control center, a turbine governor, and a DER aggregator for each load bus, as shown in Fig. 3.

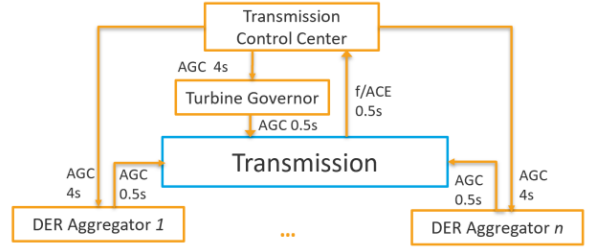

Fig. 3. Simulation components with information exchange.

The transmission dynamic simulator sends the system frequency and the ACE signals to the transmission control center, where the AGC signals are calculated with a proportional and integral controller and sent to the turbine governors and the DER aggregator. This setup is modeled in HELICS, where the transmission dynamic simulation federate uses ANDES. The setup of the federation is shown in Fig. 4.

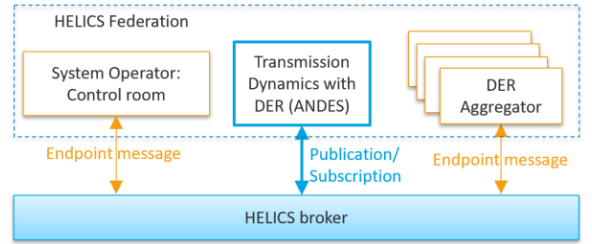

Fig. 4. HELICS federation setup.

### IV. TEST SYSTEM AND CASE STUDIES

The IEEE 39-bus system is used to evaluate the impact of the communication delay on the DER AGC signals. Assume the following system operating condition: 40 DERs at every load bus, for a total of 19 load buses with 760 DERs; the generation of the DERs is 20% of the loads at every load bus, and they are distributed evenly. The DER frequency dynamics with PFR and SFR have been added in ANDES, as described in Section II.

A generation outage is created at the 5$^{th}$ second, Fig. 5 shows the frequency behaviors of the system, and Fig. 6 shows the DER AGC signals; both figures include various delay scenarios. One can observe that the 4-s delay causes system instability, and thus the delay margin is 3 s in this setup ($k_p$=0.2, $k_i$=0.2).

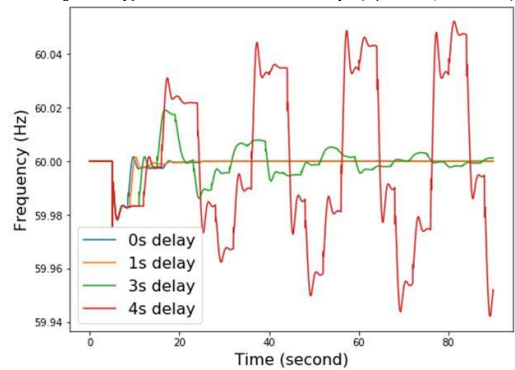

Fig. 5. System frequency response under different DER delay signals.

The delay margins for different PI controller parameters, $k_p$'s and $k_i$'s, are shown in Fig. 7 as a 3D plot; the enclosed space of the two surfaces is the feasible space of the three values ($k_p$, $k_i$, and delay), ensuring the stability of the system. Fig. 8 is the feasible space but for turbine governor AGC scenarios. A comparison of the two shows that the upper and lower delay margins (surfaces) are quite different. In the DER case, when $k_p$, $k_i$ are large, shorter delays can cause system instability, whereas longer delays do not. In the turbine governor case, however, generally longer delays tend to have a higher risk of instability. This demonstrates the importance of incorporating delay models when designing DER AGC controls.

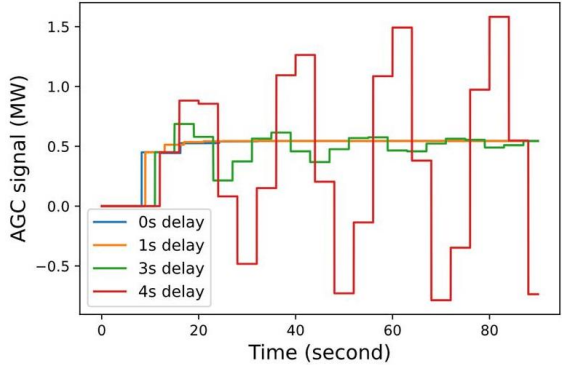

Fig. 6. DER AGC signal under different communication delays.

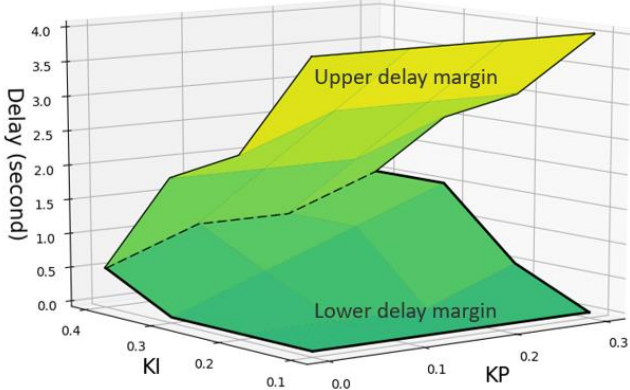

Fig. 7. Feasible space of the three values in DER AGC controls.

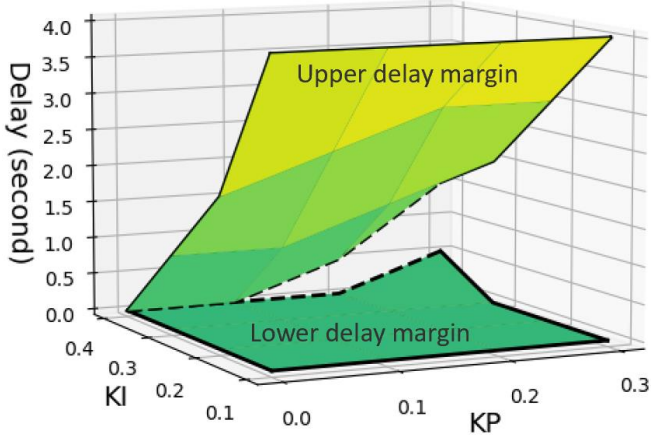

Fig. 8. Feasible space of the three values in turbine goverer AGC controls.

## V. CONCLUSION

This paper investigates the impacts of the delayed AGC signals for DERs on the frequency stability of the system. A cyber-physical dynamic simulation model is developed based on the HELICS platform. The DER frequency dynamics are modeled in the transmission simulation with delays, realized by ANDES. The simulation results with more than 700 DERs show that the risk of the system instability might be increased substantially if the design of the AGC control fails to consider communication variations. The feasible space of the communication delays in the DER AGC can be quite different from that of the conventional generators providing AGC; therefore, system operators should carefully configure the control parameters and consider the communication delays when dispatching DERs for AGC services. More theoretical analysis regarding the stability of discrete AGC signal of DERs will be our future research focus.

ACKNOWLEDGMENTS

This work was authored by Alliance for Sustainable Energy, LLC, the manager and operator of the National Renewable Energy Laboratory for the U.S. Department of Energy (DOE) under Contract No. DE-AC36-08GO28308. Funding provided by the U. S. Department of Energy Office of Electricity's Advanced Grid Research and Development program. The U.S. Government retains and the publisher, by accepting the article for publication, acknowledges that the U.S. Government retains a nonexclusive, paid-up, irrevocable, worldwide license to publish or reproduce the published form of this work, or allow others to do so, for U.S. Government purposes. The views expressed in the article do not necessarily represent the views of the DOE or the U.S. Government.